\documentclass[a4paper,12pt]{article}
\usepackage{epsfig}
\usepackage[sumlimits,intlimits,namelimits]{amsmath}

\usepackage[english]{babel}

\usepackage{graphicx}

\usepackage{geometry}
\numberwithin{equation}{section}
\usepackage{cite}

\def\slash#1{#1 \hskip-0.50em /}

\begin{document}
\begin{titlepage}
\begin{flushright}
SI-HEP-2007-06 \\[0.2cm]
September 2007
\end{flushright}

\vspace{1.2cm}
\begin{center}
{\Large\bf
Model-independent Analysis  \\[2mm]  of Lepton Flavour Violating
\boldmath $\tau$ \unboldmath Decays}
\end{center}

\vspace{0.5cm}
\begin{center}
{\sc B.~M.~Dassinger, Th.~Feldmann,
     Th.~Mannel, S.~Turczyk} \\[0.1cm]
{\sf Theoretische Physik 1, Fachbereich Physik,
Universit\"at Siegen\\ D-57068 Siegen, Germany}
\end{center}

\vspace{0.8cm}
\begin{abstract}
\vspace{0.2cm}\noindent
Many models for physics beyond the Standard Model predict lepton-flavour violating 
decays of charged leptons at a level which may become observable very soon. In the 
present paper we investigate the decays of a $\tau$ into three charged leptons in 
a generic way, based on effective-field-theory methods, where
the relevant operators are classified according to their chirality structure.
We work out the decay distributions and discuss phenomenological implications.
\end{abstract}

\end{titlepage}

%papercontent
\newpage
\pagenumbering{arabic}
\section{Introduction}

Lepton flavour violation (LFV) has become a hot topic over the last few years.
On the one hand, the discovery of neutrino oscillations in combination with the minimal extension of the Standard Model (SM) predicts lepton
flavour violation for the charged leptons, however, at a completely unobservable
level. On the other hand, many extensions of the SM predict LFV at much higher rates, which, in some cases, may already be in conflict with existing experimental bounds \cite{PDG}
(see also \cite{Banerjee:2007rj} for a recent summary of $B$\/-factory results).
With the advent of new experimental facilities 
\cite{Yamada:2005tg} (see also \cite{SuperB})
the current bounds will be pushed further, 
if not a discovery will be made. In particular, at the LHC experiments it
will be possible to detect LFV decays of a $\tau$ lepton, especially into channels
with three leptons; here the signal $\tau \to 3 \mu $ will be one of
the cleanest signatures \cite{LHCexp}.

There are many models which predict LFV $\tau$ decays of the form
$\tau \to \ell \ell' \ell '' $ with $\ell, \ell', \ell '' = e, \mu$
\cite{Ilakovac:1994kj,Barbieri:1995tw,Hisano:1995cp,Ellis:2002fe,Brignole:2004ah,Masiero:2004js,Arganda:2005ji,Paradisi:2005fk,Chen:2006hp,Goyal:2006vq,Choudhury:2006sq,BurasBlanke}. 
All these models will eventually
match onto a set of local four-fermion operators or radiative operators, the latter mediating
$\tau \to \ell \gamma^*$ with subsequent decay of the (virtual) 
photon into a charged lepton pair.
In a bottom-up approach, this allows us to
consider all possible four-fermion and radiative operators with arbitrary coupling constants, 
which can be determined by studying the decay distributions of the three leptons in the final state. Even if no signal events are found, such a study of the decay distributions
is necessary to determine the efficiency of an experiment and hence to extract reliable
limits.

Our paper is organized as follows.
In the next section we will introduce the general set of effective operators
of dimension six and eight, which mediate the decays $\tau \to \ell \ell' \ell''$ at
the electroweak scale. This includes four-fermion operators with
scalar, vector and tensor currents as well as radiative operators
contributing to $\tau \to \ell \gamma \to \ell \ell' \ell''$. 
These operators match onto the relevant low-energy interactions at the scale 
of the $\tau$ mass, which are parameterized by a number of unknown coupling constants. 
We will also give a brief discussion on how
the couplings are related to the lepton masses and the PMNS neutrino mixing matrix
within minimal-flavour violating scenarios \cite{Cirigliano:2005ck}.
Focusing on the decays $\tau^- \to \mu^-\mu^- \mu^+$ in section~\ref{sec:taumu}
and $\tau^- \to e^-\mu^- \mu^+$ in section~\ref{sec:taue},
we calculate the Dalitz distributions for the individual chirality structures
appearing in the effective Hamiltonian, taking into account interference
terms apart from corrections of order $m_\mu/m_\tau$.
We conclude with a brief comparison of existing results on LFV processes
in specific new physics scenarios in section~\ref{sec:concl}.

%%%%%%%%%%%%%%%%%%%%%%%%%%%%%%%%%%%%%%%%%%%%%%%%%%%%%%%%%%%%%%%%%%%%%%%%%%%%%%%%%%%%%%%%%%

\section{The Effective Interaction for $\tau \to \ell \ell' \ell''$ Decays}

In this paper we follow a bottom-up approach to 
lepton-flavour violating decays, using an effective-field-theory
picture. We assume that some new physics
at a high scale $\Lambda$ induces lepton-flavour violating processes.\footnote{Notice that,
in general, the scale associated to lepton-flavour violation is independent
of the scale related to lepton-number violation, $\Lambda_{\rm LN}$.}
At the electro-weak scale these
lepton-flavour-violating interactions manifest themselves in 
higher dimensional operators
which have to be compatible with the $SU(2)_L \times U(1)_Y$ gauge symmetry of the SM. 
The leading operators for processes involving charged leptons will be of dim=6 and dim=8, 
which we shall construct in the following (similar analyses within the context of
supersymmetric extensions of the SM can be found, for instance, in
\cite{Hisano:1995cp,Brignole:2004ah}).

To this end we group the left-handed leptons in an $SU(2)_L$
doublet, while the right-handed charged leptons (which are singlets
under $SU(2)_L$) are put into an incomplete doublet,
as a reminiscent of a right-handed $SU(2)_R$ related to custodial symmetry.
Writing also the Higgs boson in matrix form, we have
\begin{equation} \label{fields}
L =  \left(\begin{array}{c} \nu_L \\ \ell_L \end{array} \right) \, ,  \quad
R = \left(\begin{array}{c} 0 \\ \ell_R \end{array} \right)  \, , \quad
H = \frac{1}{\sqrt{2}}
    \left( \begin{array}{cc}
           v+h_0 + i \chi_0   &\   \sqrt{2} \phi_+ \\
           - \sqrt{2} \phi_-   &\   v+h_0 - i \chi_0 
                                                       \end{array} \right)
\,.
\end{equation} 
For simplicity, we have suppressed the family indices, which will be specified once we consider a particular decay mode.
For later use, we note that the 
hypercharge $Y$ is defined in terms of the right-handed generator $T_3^{(R)}$ as
\begin{equation} \label{hyper} 
Y = T_3^{(R)} + \frac{1}{2} (B-L) \,.
\end{equation} 
In terms of the fields defined in (\ref{fields}),
the list of operators relevant for $\tau \to \ell\ell'\ell''$ decays
reads
\begin{center}
\begin{minipage}{8cm}
\begin{align}
& \mbox{\textbf{dim=6 leptonic:}}\nonumber \\
& O_1 = (\bar{L} \gamma_\mu L)  (\bar{L} \gamma^\mu L)  \label{O1}\\
& O_2 = (\bar{L} \tau^a  \gamma_\mu L)  (\bar{L} \tau^a \gamma^\mu L)  \\
& O_3 =  (\bar{R} \gamma_\mu R)  (\bar{R} \gamma^\mu R) \\
& O_4 =  (\bar{R} \gamma_\mu R)  (\bar{L} \gamma^\mu L) \label{O4}\\
& \quad \nonumber  \\
& \mbox{\textbf{dim=6 radiative:}} \nonumber \\
& R_1 =  g' (\bar{L} H  \sigma_{\mu \nu}  R) B^{\mu \nu}  \label{R1}\\
& R_2 =  g (\bar{L}  \tau^a H \sigma_{\mu \nu}   R) W^{\mu \nu, a} \label{R2}\\
& \quad \nonumber
\end{align}
\end{minipage}
\hspace{0.5cm}
\begin{minipage}{8cm}
\begin{align}
& \mbox{\textbf{dim=8 leptonic:}}\nonumber \\
& P_1 =  (\bar{L} H  R)  (\bar{L} H R) \label{P1}\\
& P_2 =  (\bar{L} \tau^a H  R)  (\bar{L} \tau^a H R) \\
& Q_1 =  (\bar{L} H  R)  (\bar{R} H^\dagger L) \\
& Q_2 =  (\bar{L} \tau^a H  R)  (\bar{R} H^\dagger \tau^a L) \\
& P_1^{(T)}  =  (\bar{L} H  \sigma_{\mu \nu} R)  (\bar{L} H \sigma^{\mu \nu} R) \\
& P_2 ^{(T)} =  (\bar{L} \tau^a H \sigma_{\mu \nu} R)  (\bar{L} \tau^a H \sigma^{\mu \nu} R) 
\end{align}
\end{minipage}
\end{center}
where $B_{\mu \nu}$ is the $U(1)_Y$ gauge field, $W_{\mu \nu}^a$ are the
$SU(2)_L$ gauge fields, and $g$ and $g'$ are the corresponding gauge couplings.
In the above list we have only shown operators that have tree-level contributions
to leptonic $\tau$\/-decays; more operators, which are bi-linear in the lepton fields
and contribute at the loop level, can be found e.g.\ in \cite{Hisano:1995cp,Brignole:2004ah,Cirigliano:2005ck}.
We also neglected dim=8 operators involving additional covariant derivatives. 
When acting on fermions, the derivatives become fermion masses by the equations of motions, 
such that these operators are additionally suppressed by the small lepton Yukawa couplings.
Notice that the leptonic dim=6 operators $O_1-O_4$ only contain helicity
conserving currents. 
The most general effective Hamiltonian at the electro-weak scale
is then obtained by summing over these operators, multiplied by arbitrary 
coefficients for every flavour combination. 
In a particular new physics scenario, these coefficient should be obtained by matching 
at the new physics scale $\Lambda$ and evolving down to the scale $M_W$ within the
SM (as an effective theory).

In the following we are interested in LFV decays of a $\tau$ lepton into three charged
leptons. To this end we have to construct the effective interaction at the scale of the
$\tau$ lepton, by integrating out the weak gauge bosons and the Higgs. 
We will focus on $\tau^-$ decays; the decay distributions for $\tau^+$ decays are 
identical. On the level of four-fermion operators with dim=6, we obtain the same
structures as in (\ref{O1}-\ref{O4}). Projecting on charged
leptons only, we see that $O_2$ becomes equivalent to $O_1$, and both
match onto a purely left-handed operator
\begin{equation} \label{LLLL}
H_{\rm eff}^{(LL)(LL)} =
g_V^{(LL)(LL)} \, \frac{(\bar{\ell}_L \gamma_\mu \tau_L )(\bar{\ell}'_L \gamma^\mu \ell''_L)}{\Lambda^2} \,,
\end{equation}
where here and in what follows the superscript of the coupling denotes the
combinations of chiralities involved and the subscript denotes the relevant
Dirac structure.
Likewise, the operator $O_3$ corresponds to a purely right-handed interaction
\begin{equation} \label{RRRR}
H_{\rm eff}^{(RR)(RR)} =
g_V^{(RR)(RR)} \, \frac{(\bar{\ell}_R \gamma_\mu \tau_R )(\bar{\ell}'_R \gamma^\mu \ell''_R)}{\Lambda^2} \,,
\end{equation}
while we get a mixed term from the operator $O_4$
\begin{equation} \label{LLRR}
H_{\rm eff}^{(LL)(RR)} =
g_V^{(LL)(RR)} \, \frac{(\bar{\ell}_L \gamma_\mu \tau_L )(\bar{\ell}'_R \gamma^\mu \ell''_R)}{\Lambda^2}
+ g_V^{(RR)(LL)} \, \frac{(\bar{\ell}_R \gamma_\mu \tau_R )(\bar{\ell}'_L \gamma^\mu \ell''_L)}{\Lambda^2} \,.
\end{equation}
Notice that the chirality structure $(LR)(RL)$ is not independent, since it
can be Fierz rearranged into $H_{\rm eff}^{(LL)(RR)}$. % (see appendix).

The dim=6 radiative operators contain charged as well as neutral currents.
Here we are  only interested in the neutral-current component, 
coupling to a charged lepton pair.
Switching to the physical photon and $Z_0$ fields and integrating out the Higgs and the $Z_0$,
we find that we obtain a radiative operator with a photon, as well as
a four-fermion contribution from $Z_0$ exchange. The latter is proportional to
\[
\frac{v}{\Lambda^2} \frac{1}{v^2}  (\bar{\ell}  \sigma_{\mu \nu}   \tau) q^\nu
 (\bar{\ell}'  \gamma^\mu (g_V + g_A \gamma_5) \ell'' ) \, ,
\]
where $(g_A) \, g_V$ are the (axial)vector couplings of the $Z_0$ to the leptons.
Taking into account the fact that $|q^\mu|$ is of the order of the $\tau$ mass, we find
that this operator is suppressed relative to the leading ones by the small Yukawa
coupling of the $\tau$ lepton.  Thus, only the photonic contribution has to be taken
into account. For this we obtain at the scale $m_\tau$
\begin{equation} \label{rad}
H_{\rm eff}^{\rm rad} =
\frac{e}{4 \pi} \, \frac{v}{\Lambda^2} \, \sum_{h,s}  g_{\rm rad}^{(s,h)}
\left(\bar{\ell}_h (-i \sigma_{\mu\nu})  \tau_s \right) F^{\mu \nu} \,,
\end{equation}
where $g_{\rm rad}^{(L,R)}$ and $g_{\rm rad}^{(R,L)}$ 
denote the two possible chirality combinations.\footnote{
Here, for simplicity, we neglected possible form factor effects  for
decays into virtual photons from long-distance lepton or quark loops. In the
most general case, the $\tau \to \ell \gamma^*$ vertex could be parametrized as
\[
\frac{e}{4 \pi} \, \frac{v}{\Lambda^2} \, \sum_{h,s} 
\, \bar{\ell}_h \left\{
 g_{\rm rad}^{(s,h)}(q^2) \, (-i \sigma_{\mu\nu}) \, q^\mu
+ m_\tau \, f_{\rm rad}^{(s,h)}(q^2) \, \left(\gamma_\nu - \frac{q_\nu}{q^2} \, \slash q \right)
\right\}
 \tau_s \,,
\]
where $g_{\rm rad}^{(s,h)}(0) \equiv g_{\rm rad}^{(s,h)}$ 
and $f_{\rm rad}^{(s,h)}(0)=0$, see, for instance, \cite{Raidal:1997hq}.
}
The matrix element for $\tau \to \ell  \bar{\ell}' \ell'$ becomes
\begin{equation}\label{radhad}
\langle \ell \bar{\ell}' \ell' | H_{\rm eff}^{\rm rad} | \tau \rangle =
\alpha_{em}   \,  \frac{v}{\Lambda^2} \, \frac{q^\nu}{q^2}\, \sum_{h,s}  g_{\rm rad}^{(s,h)}
 \, \langle \ell \bar{\ell}' \ell' |
\left(\bar{\ell}_h (-i \sigma_{\mu\nu})  \tau_s\right) \left(\bar{\ell}' \gamma^\mu \ell'\right)| \tau \rangle \,,
\end{equation}
where $q$ is the momentum transfer through the photon. This momentum transfer
is proportional to the lepton masses, and thus this contribution scales as
$1/(y \Lambda^2)$ where $y$ is a Yukawa coupling of the leptons, which would
lead to an enhancement unless an additional Yukawa coupling appears in the
numerator as e.g. in minimal flavour violation (see below).

Finally, we turn to the four-fermion operators with the chirality structure
$(RL)(RL)$ or $(LR)(LR)$. At tree-level, they receive contributions from
the dim=8 operators $P_{1,2}$, $Q_{1,2}$
and $P_{1,2}^{(T)}$, only. Therefore, their matching coefficients are 
further suppressed by $v^2/\Lambda^2$. 
The one-loop matching coefficients at the electroweak scale may
also receive contributions from the dim=6 operators $O_{1-4}$, $R_{1,2}$,
but in this case the required chirality flips induce an additional suppression 
by $m_\ell^2/v^2$.
Ignoring contributions suppressed by $v^2/\Lambda^2$ or $m_\ell^2/v^2$,
this reduces the number of possible Dirac structures already 
to six in the case where the radiative operator can contribute and to four in cases
like $\tau^- \to \mu^- \mu^- e^+$, where the radiative contribution is absent. 
The corresponding couplings 
\[
g_V^{(LL)(LL)}, \quad g_V^{(RR)(RR)}, \quad  g_V^{(LL)(RR)}, \quad
g_V^{(RR)(LL)}, \quad g_{\rm rad}^{(LR)}, \quad g_{\rm rad}^{(RL)} ,
\]
are matrices in lepton flavour space.
There are in total six different decay modes of the $\tau^-$ to consider,
\begin{center}
\begin{minipage}{7cm}
\begin{align}
\tau^- &\to  e^- e^- e^+\label{decay1}\\
\tau^- &\to \mu^-  \mu^- \mu^+ \label{decay2}\\
\tau^- &\to  e^- e^- \mu^+\label{decay3}
\end{align}
\end{minipage}
\hspace{1cm}
\begin{minipage}{7cm}
\begin{align}
\tau^- &\to \mu^-  \mu^-  e^+\label{decay4}\\
\tau^- &\to  \mu^- e^-  e^+\label{decay5}\\
\tau^- &\to e^-  \mu^- \mu^+\label{decay6}
\end{align}
\end{minipage}
\end{center}
Notice that (\ref{decay1} - \ref{decay4}) contain two identical
particles ($e^-e^-$ or $\mu^-\mu^-$) in the final state, whereas
(\ref{decay5} + \ref{decay6}) do not.
Moreover, only (\ref{decay1}, \ref{decay2}, \ref{decay5}, \ref{decay6})
receive contributions from the radiative operators
(\ref{rad}) via
\[
 \tau^- \to \ell^- \gamma^* \to \ell^- (\ell'{}^+ \ell'{}^-) \,.
\]

%%%%%%%%%%%%%%%%%%%%%%%%%%%%%%%%%%%%%%%%%%%%%%%%%%%%%%%%%%%%%%%%%%%%%%%%%%%%%%%%%%%%%%

\subsection{Constraints from Minimal Flavour Violation}

%%%%%%%%%%%%%%%%%%%%%%%%%%%%%%%%%%%%%%%%%%%%%%%%%%%%%%%%%%%%%%%%%%%%%%%%%%%%%%%%%%%%%%

The flavour structure of the coupling constant has been investigated within
the framework of Minimal Flavour Violation in the lepton sector 
(MLFV \cite{Cirigliano:2005ck}). 
In a scenario with the minimal field content (\ref{fields}), the breaking
of the lepton flavour symmetry $SU(3)_L\times SU(3)_{E_R}$
is described by two spurion fields
\begin{eqnarray}
\lambda &=& \frac{m_\ell}{v} = \frac{1}{v} \, {\rm diag} (m_e, m_\mu, m_\tau) \,, \\
g_\nu &=& \frac{\Lambda_{\rm LN}}{v^2} \, 
U^* \, {\rm diag} (m_{\nu_1}, m_{\nu_2}, m_{\nu_2}) \, U^\dagger 
\end{eqnarray}
where $\lambda \sim (\bar 3,3)$ describes the SM Yukawa couplings of the charged leptons,
and the matrix $g_\nu \sim (\bar 6,1)$ stems from a dim-5 lepton-number violating term,
\begin{equation} \label{Maj}
{\cal L}_{\rm Maj} = \frac{1}{2 \Lambda_{\rm LN}} \left( N^T g N \right)\,,
\end{equation}
where \begin{equation}
N = \left(T_3^{(R)} + \frac{1}{2} \right) H^\dagger L   
\end{equation}
has vanishing quantum numbers under the complete SM gauge group.
If the scale $\Lambda_{\rm LN}$, associated with lepton-number violation,
is sufficiently large, the resulting neutrino masses
$m_{\rm Maj} \sim v^2/\Lambda_{\rm LN}$ are small, 
even if the spurion $g_\nu$ has generic entries of order unity.

We are interested in 4-lepton processes, induced by operators with some flavour
structure
\[
  L^i \, L^j \, L_k^* \, L_l^* \,, \quad
  L^i \, R^j \, L_k^* \, R_l^* \,, \quad \mbox{etc.}
\]
To render these operators formally invariant under the flavour group, they have
to be multiplied by appropriate factors of $\lambda_e$ and $g_\nu$. In the 
following we will consider the minimal number of spurion insertions, only.\footnote{
This is justified as long as the spurion fields are characterized by some
small expansion parameter \cite{Cirigliano:2005ck,D'Ambrosio:2002ex,Feldmann:2006jk}, e.g.\ if the neutrino mass
{\em differences} $\Delta m_\nu^2$ are smaller than their average $\Delta\bar m_\nu^2$.
Notice that, unlike in the case of the quark CKM matrix, the off-diagonal entries
of the PMNS matrix are not always small.}

Starting with the case of $L^i \, L^j \, L_k^* \, L_l^*$,
we need at least two spurion insertions. The possible flavour structures 
can be read off the reduction of the $SU(3)_L$ tensor product
for $g_\nu$ and $g_\nu^\dagger$,
$$
  \bar 6 \times 6 = 1 + 8 + 27 \,.
$$
Here the flavour singlet term corresponds to the trace of $g_\nu^\dagger g_\nu$,
$$
  {\rm tr}[g_\nu^\dagger g_\nu] = \frac{\Lambda_{\rm LN}^2}{v^4} 
      \left( m_{\nu_1}^2+m_{\nu_3}^2+m_{\nu_3}^2\right) 
\equiv 3 \, \frac{\Lambda_{\rm LN}^2 \, \bar m_\nu^2}{v^4} \,,
$$
which does not induce flavour transitions at all.
The flavour octet term is obtained as\footnote{Notice 
that our definition of $\Delta$ differs from the one in \cite{Cirigliano:2005ck},
but only for the diagonal elements, which are irrelevant for flavour transitions.}
\begin{eqnarray}
 \Delta =\Delta^\dagger &=& g_\nu^\dagger g_\nu - \frac{1}{3} \, {\rm tr}[g_\nu^\dagger g_\nu] 
 = \frac{\Lambda_{\rm LN}^2}{v^4} \, U \, \Delta m_\nu^2 \,  U^\dagger\,.
\end{eqnarray}
Here $\Delta m_\nu^2 = {\rm diag}[m_{\nu_1}^2,m_{\nu_2}^2, m_{\nu_3}^2]- \bar m_\nu^2$.
In particular, one finds that 
$$
  \Delta_\tau^\mu = {\cal O}\left( \frac{\Lambda_{LN}^2}{v^4} \, \Delta m_{\rm atm}^2 \right)
$$
whereas $\Delta_\mu^e$ and $\Delta_\tau^e$ are further suppressed by the neutrino
mixing angle $\theta_{13}$.
It is to be stressed that $\Delta$ does neither depend on the absolute neutrino mass scale 
$\bar m_\nu^2$, nor on potential Majorana phases $\alpha_{1,2}$ in the PMNS matrix.

The flavour structure of the corresponding invariant 4-lepton operator reads
$
  (L^* \, \Delta \, L)(L^* \, L) \,.
$
The coefficients of specific flavour transitions are thus given by the quadratic
neutrino mass differences and PMNS elements. 
By the same argument, the operator
$
   (L^* \, \Delta \, L)(R^* \, R)  
$
is invariant under the flavour group.
It has been shown in \cite{Cirigliano:2005ck} that $\Delta$ also drives 
all possible flavour transitions induced by operators that are 
bilinear in the lepton fields. In particular, the flavour structure of
the radiative operators in (\ref{R1},\ref{R2}) reads
$
 (L^* \, \Delta \, \lambda^\dagger R)$ and $ (R^* \lambda \, \Delta \, L) \,
$.
Notice that the presence of a single right-handed field requires the insertion
of the Yukawa spurion $\lambda$, which leads to an additional suppression factor
$m_\ell/v$.

Turning to the 27plet combination of $g_\nu$ and $g_\nu^\dagger$, 
we introduce the according representation in terms of a trace-less tensor
\begin{eqnarray}
 G_{ij}^{kl} &=& (g_\nu)_{ij} \, (g_\nu^*)^{kl} 
  - \frac{1}{12} \left(\delta_i^k \delta_j^l + \delta_i^l \delta_j^k \right) 
     {\rm tr}(g^\dagger g) \nonumber\\[0.2em]
&& {} 
  - \frac{1}{5} \left( 
      \delta_i^a \delta_b^l \delta_j^k + \delta_j^a \delta_b^l \delta_i^k
    + \delta_i^a \delta_b^k \delta_j^l + \delta_j^a \delta_b^k \delta_i^l
   \right) \, \Delta_a^b
\end{eqnarray}
with $G_{ij}^{kl}=  G_{ji}^{kl}= G_{ij}^{lk}$, and
$
  \sum_i \, G_{ij}^{il} = 0
$.
The flavour structure of the corresponding invariant 4-lepton operator reads
$$
  G_{ij}^{kl} \, L^i L^j \, L^*_k L^*_l \,.
$$
In contrast to $\Delta$, the off-diagonal matrix elements of $G$ depend on the
absolute neutrino mass scale $\bar m_\nu^2$ and the Majorana phases. As a
consequence, in the
general case, i.e.\ if the radiative operators do not dominate the $\tau \to 3\ell$
decay amplitudes, the purely leptonic decay modes are not directly correlated with
the radiative ones $\tau \to \ell\gamma$. 
Relatively simple expressions for $G_{ij}^{kl}$ can be obtained in the
limit of vanishing Majorana phases, where we also employ the approximations
$\sin^2 \theta_{13} \sim \Delta m^2_{\rm sol}/\Delta m^2_{\rm atm} \ll 1$
and $\theta_{23}=45^\circ$.
For the normal neutrino hierarchy ($m_{\nu_1} \sim m_{\nu_2} \ll m_{\nu_3}$),
we obtain the leading coefficients as
\begin{equation}
 G_{\tau e}^{e\mu} \simeq - 2 G_{\tau \mu}^{\mu\mu} \simeq 
 - \frac{\Lambda_{\rm LN}^2}{v^4} \, \frac{\Delta m^2_{\rm atm}}{10} \,, 
\end{equation}
and sub-leading effects from
\begin{align}
& G_{\tau e}^{ee} \simeq - \frac{\Lambda_{\rm LN}^2}{v^4} \,\frac{\sqrt2}{5} \, e^{i\delta} \, \sin\theta_{13} \, \Delta m^2_{\rm atm} 
 \,, \qquad
 G_{\tau\mu}^{ee} \simeq\frac{\Lambda_{\rm LN}^2}{v^4} \,
\frac{\sqrt{  \Delta m^2_{\rm atm}}}{2}
\left( m_{\nu_{1,2}} - \cos2\theta_{12} \, \frac{\Delta m^2_{\rm sol}}{2m_{\nu_{1,2}}} \right)
 \,, \nonumber\\
& G_{\tau e}^{\mu\mu} \simeq \frac{\Lambda_{\rm LN}^2}{v^4} \,\frac{\sqrt{  \Delta m^2_{\rm atm}}}{2\sqrt2} \left(
   e^{i\delta} \, \sin\theta_{13} \, \sqrt{  \Delta m^2_{\rm atm}} 
  - \sin2\theta_{12} \, \frac{\Delta m^2_{\rm sol}}{4m_{\nu_{1,2}}}
\right)
\,, \nonumber \\ 
& G_{\tau\mu}^{e\mu} \simeq  \frac{\Lambda_{\rm LN}^2}{v^4} \,
\frac{\sqrt{  \Delta m^2_{\rm atm}}}{2}
\left( \frac{3 \cos\delta -7 i \sin \delta}{5} \, \sin\theta_{13} \, \sqrt{  \Delta m^2_{\rm atm}}
+\sin2\theta_{12} \, \frac{\Delta m^2_{\rm sol}}{4m_{\nu_{1,2}}}
\right) \,.
\end{align}
where we have used the PDG parameterization \cite{PDG} of the CKM matrix.
For the inverted hierarchy ($m_{\nu_1} \sim m_{\nu_2} \gg m_{\nu_3}$), one has
\begin{equation}
 G_{\tau\mu}^{ee} \simeq 
  -5 G_{\tau e}^{e\mu} \simeq 10 G_{\tau \mu}^{\mu\mu} 
 \simeq- \frac{\Lambda_{\rm LN}^2}{v^4} \,\frac{\Delta m^2_{\rm atm}}{2}
 \,, 
\end{equation}
and
\begin{align}
& G_{\tau e}^{ee} \simeq -\frac{\Lambda_{\rm LN}^2}{v^4} \,
    \frac{\Delta m^2_{\rm atm}}{\sqrt2} \, \sin\theta_{13} 
  \, \frac{3 \cos\delta - 7 i \sin\delta}{5}
 \,, \nonumber\\
& G_{\tau e}^{\mu\mu} \simeq -\frac{\Lambda_{\rm LN}^2}{v^4} \,
  \frac{\Delta m^2_{\rm atm}}{2 \sqrt2} \, e^{i\delta} \, \sin\theta_{13}
\,, \qquad
 G_{\tau\mu}^{e\mu} \simeq \frac{\Lambda_{\rm LN}^2}{v^4} \,
\frac{7 \Delta m^2_{\rm atm}}{10 \sqrt2} \, e^{i\delta} \, \sin\theta_{13}
\,.
\end{align}

We finally note that purely right-handed lepton-flavour
violating decays require at least four spurion insertions,
$
  (R^* R) (R^*  \lambda \, g^\dagger  g \, \lambda^\dagger  R) \,,
$
and are thus strongly suppressed in MLFV.

In summary, to obtain the dominating flavour coefficients
for the operators in
(\ref{LLLL}, \ref{LLRR}, \ref{rad}),
relevant for flavour-violating $\tau$ decays in MLFV, 
one has to consider 
\begin{eqnarray}
g_V^{(L_k L^i)(L_l L^j)} &\to&
2 c_1 \, \Delta^{k}_i \, \delta_{j}^{l}  
+ c_2 \,  G_{ij}^{kl}   \,,
 \\[0.1em]
g_V^{(L_k L^i)(R_l R^j)}
 &\to &
c_3 \, \Delta_i^k \,\delta_{j}^{l} \,,
\\[0.1em]
g_{\rm rad}^{(L_k R^i)} &\to & c_4 \, \Delta^k_i 
\,,
\end{eqnarray}
whereas the chiral structures corresponding to $g_V^{(RR)(RR)}$, $g_V^{(RR)(LL)}$
and $g_{\rm rad}^{(RL)}$ are suppressed by small lepton masses.
The spurion combination $G_{ij}^{kl}$ represents
a new source of LFV compared to the radiative transitions $\tau \to \ell \gamma$. 
While the latter are driven by the spurion $\Delta$ and hence by the difference 
of the squared neutrino masses, the flavour coefficients of
purely left-handed four-lepton operators in MLFV also involve 
the absolute neutrino mass scale as well as the Majorana phases. 
In particular, the decay modes (\ref{decay3} + \ref{decay4}) only receive
contributions from $G_{\tau \mu}^{ee}$ and $G_{\tau e}^{\mu \mu}$, 
respectively.

%%%%%%%%%%%%%%%%%%%%%%%%%%%%%%%%%%%%%%%%%%%%%%%%%%%%%%%%%%%%%%%%%%%%%%%%%%%%%%%%%%%%
\begin{figure}[t!!]
      \begin{minipage}{0.48\textwidth}
       \centering
       \includegraphics[width=0.8\textwidth]{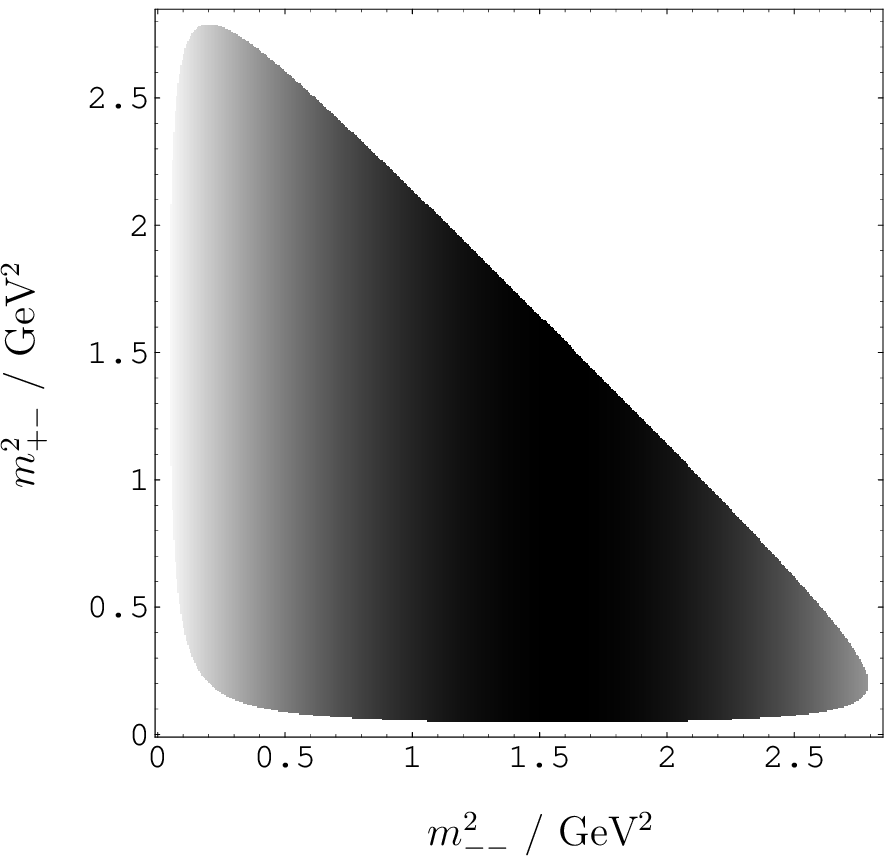}
      \end{minipage}\hfill
      \begin{minipage}{0.48\textwidth}
       \centering
        \includegraphics[width=0.8\textwidth]{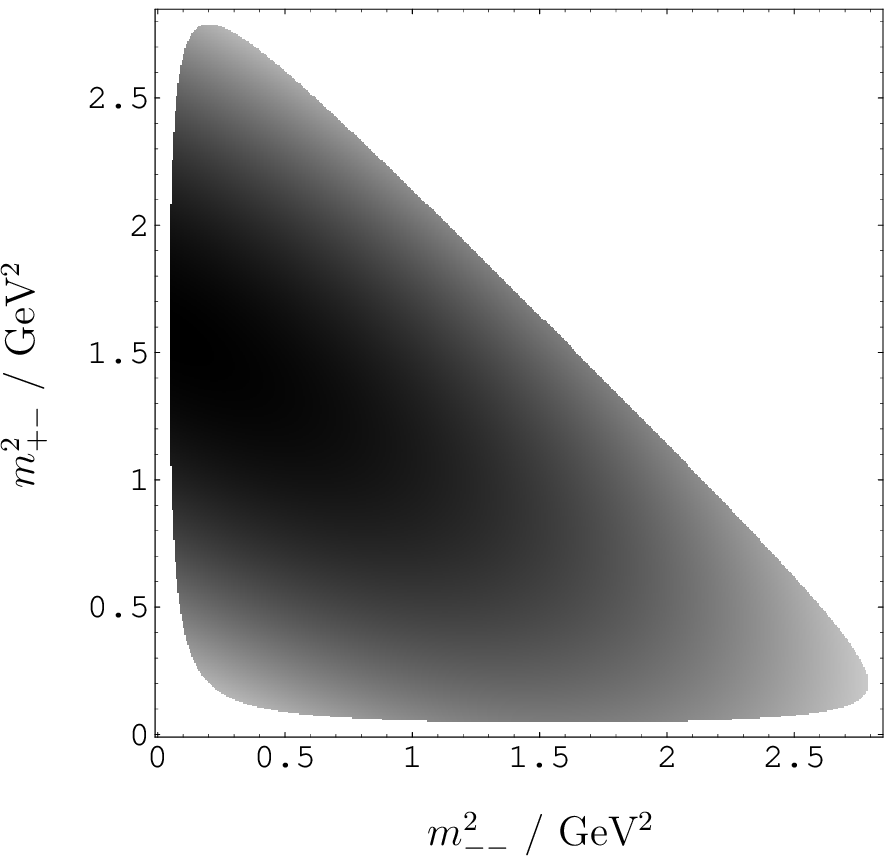}
      \end{minipage}
  \caption{Dalitz plot for $\text{d}^2\Gamma_V^{(LL)(LL)}$ (left)
      and $\text{d}^2\Gamma_V^{(LL)(RR)}$ (right)
      in $\tau^- \to \mu^- \mu^- \mu^+$. \label{plot}}
\end{figure}
%%%%%%%%%%%%%%%%%%%%%%%%%%%%%%%%%%%%%%%%%%%%%%%%%%%%%%%%%%%%%%%%%%%%%%%%%%%%%%%%%%%%%

\section{Dalitz-Plot Analysis}

\subsection{The Decay \boldmath $\tau^- \to \mu^- \mu^- \mu^+$ \boldmath}

\label{sec:taumu}

In this section we will give a detailed analysis of 
$\tau^- \to \mu^- \mu^- \mu^+$- as the probably most prominent
channel to be looked for at the LHC.
To this end, we  shall consider the Dalitz distributions 
for the different chirality structures (\ref{LLLL}, \ref{RRRR}, \ref{LLRR}, \ref{rad}) 
in the dim=6 effective Hamiltonian, in terms of the variables
\begin{equation}
m_{--}^2 \equiv m_{12}^2 = (p_{\mu^-} + p'_{\mu^-})^2 \,, \qquad
m_{+-}^2 \equiv m_{23}^2 = (p_{\mu^-}' + p_{\mu^+})^2 \,, 
\end{equation}
and $m_{13}^2 = m_\tau^2+3m_\mu^2 - m_{--}^2-m_{+-}^2$.
We will make use of (approximate) helicity conservation, which implies
that many of the interference terms between the operators with different
chiralities are suppressed by powers of $m_\mu$, and can be ignored
to first approximation.

In the simplest case all four leptons are left-handed, and
the decay amplitude is determined by $H_{\rm eff}^{(LL)(LL)}$ in (\ref{LLLL}).
The corresponding Dalitz distribution 
\begin{equation}
\frac{\text{d}^2\Gamma_V^{(L L)(L L)}}{\text{d}m_{23}^2 \, \text{d}m_{12}^2} 
= 
\frac{| g_V^{(L_\mu L^\tau)(L_\mu L^\mu )}|^2}{\Lambda^4} \, 
\frac{(m_\tau^2-m_\mu^2)^2 - (2 m_{12}^2 -m_\tau^2-3m_\mu^2)^2}{256 \, \pi ^3\, m_{\tau }^3}
\,,
\label{dGam1}
\end{equation}
is shown in Fig.~\ref{plot} (left).
The events are equally distributed along $m_{+-}^2$, while
there is a rather flat maximum at $m_{--}^2=m_{12}^2\simeq m_\tau^2/2$.
The case with all particles right-handed is completely analogous
and yields the same distribution with $g_V^{(LL)(LL)} \to g_V^{(RR)(RR)}$.
(We remind the reader that $g_V^{(RR)(RR)}$ is expected to
be strongly suppressed within MLFV scenarios.)

Next we will consider the operator $H_{\rm eff}^{(LL)(RR)}$  in (\ref{LLRR}). 
For a left-handed $\tau$\/-lepton we obtain the Dalitz distribution
\begin{equation}
\begin{split}
 \frac{\text{d}^2\Gamma_V^{(LL)(RR)}}{\text{d}m_{23}^2 \text{d}m_{12}^2} 
= \ & 
\frac{| g_V^{(L_\mu L^\tau)(R_\mu R^\mu)}|^2}{\Lambda^4}
\left[
  \frac{(m_\tau^2-m_\mu^2)^2-4 m_\mu^2 \, (m_\tau^2+m_\mu^2-m_{12}^2)}{512 \, \pi^3 \, m_\tau^3}
\right. \\[0.2em]
& \qquad {} \left.
 - \frac{(2 m_{13}^2 -m_\tau^2-3m_\mu^2)^2+(2 m_{23}^2 - m_\tau^2-3m_\mu^2)^2}
        {1024 \,\pi^3 \, m_\tau^3}
\right]
\,,
\end{split}
\label{dGam2}
\end{equation}
shown in Fig.~\ref{plot} (right). 
In this case the events are distributed around a flat maximum at 
$m_{+-}^2\simeq m_\tau^2/2$ and $m_{--}^2\simeq 0$.
Again, the case of a right-handed $\tau$ yields the same distribution.
As pointed out above, the interference terms between (\ref{LLLL}) and 
(\ref{LLRR}) are suppressed by $m_\mu^2 / m_\tau^2$.

\begin{figure}[t!!b!ph]
       \centering
        \includegraphics[width=0.4\textwidth]{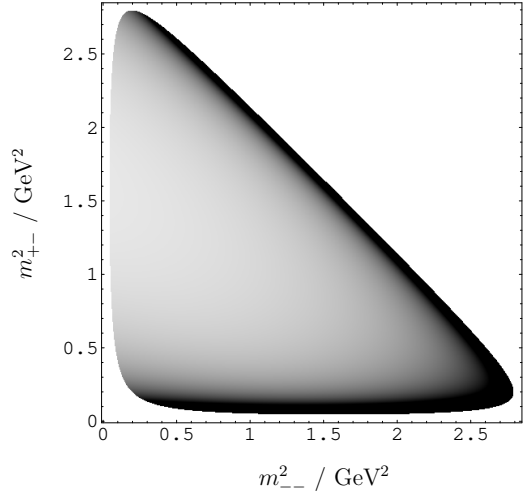}
        \caption{Dalitz plot for $\text{d}^2\Gamma_{\rm rad}^{(LR)}$ 
      in $\tau^- \to \mu^- \mu^- \mu^+$. \label{plotrad}}
\end{figure}

In addition to the four-fermion operators we also get a contribution from the 
radiative interaction via (\ref{radhad}). 
The resulting Dalitz distribution for a right-handed $\tau$\/-lepton,
\begin{equation}
\begin{split}
\frac{\text{d}^2\Gamma_{\rm rad}^{(L R)}}{\text{d}m_{23}^2 \text{d}m_{12}^2} 
=
& \
\alpha_{\text{em}}^2 \, 
\frac{|g_{\rm rad}^{(L_\mu R^\tau)}|^2 \, v^2}{\Lambda^4}
 \left[
\frac{m_\mu^2 \, (m_\tau^2 - m_\mu^2)^2}{128 \, \pi^3 \, m_{\tau }^3}
\left( \frac{1}{m_{13}^4} + \frac{1}{m_{23}^4} \right)
+ \frac{m_\mu^2 (m_\tau^4 - 3 m_\tau^2 m_\mu^2+2 m_\mu^4)}{128 \, \pi^3 \,m_{13}^2 \, m_{23}^2 \, m_{\tau }^3}
\right.\\
& \ {}
+\left. \frac{
(m_{13}^2+m_{23}^2)(m_{12}^4+m_{13}^4+m_{23}^4-6 m_\mu^2(m_\mu^2+m_\tau^2))}
{256 \, \pi^3 \,m_{13}^2 \, m_{23}^2 \, m_{\tau }^3}
+
\frac{2m_{12}^2- 3 m_\mu^2}{128 \, \pi^3  \, m_{\tau }^3}
\right]
\,,
\end{split}
\label{dGam3}
\end{equation}
is plotted in Fig.~\ref{plotrad}. 
Due to the photon pole, the events are concentrated at low values of
$m_{23}^2$ or $m_{13}^2$, respectively.
Again the decay of the left-handed $\tau$ is completely analogous.

\begin{figure}[t!pb]
      \begin{minipage}{0.48\textwidth}
       \centering
        \includegraphics[width=0.78\textwidth]{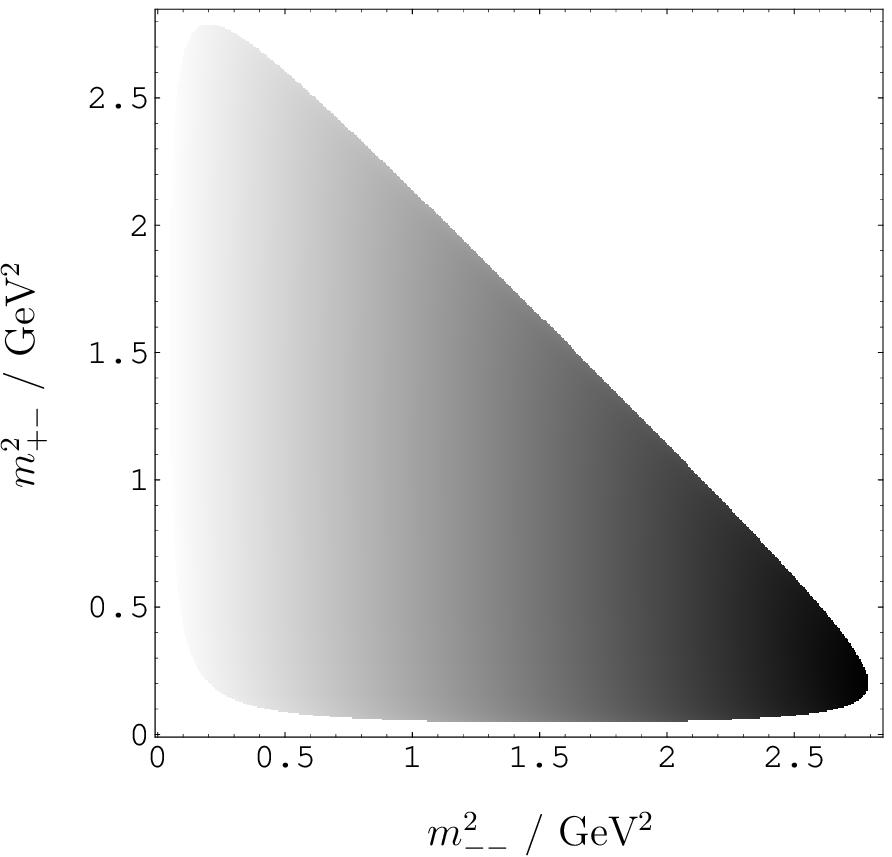}
      \end{minipage}\hfill
      \begin{minipage}{0.48\textwidth}
       \centering
        \includegraphics[width=0.78\textwidth]{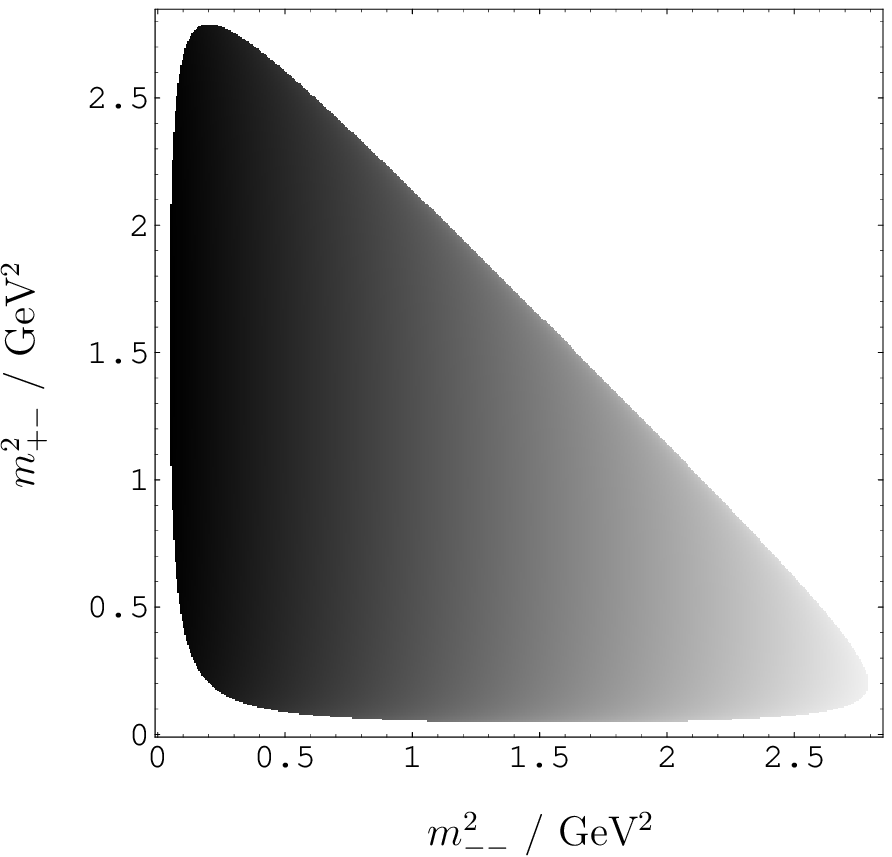}
      \end{minipage}
\caption{Dalitz plot for $|\text{d}^2\Gamma_{\rm mix}^{(LL)(LL)}|$ (left)
and $|\text{d}^2\Gamma_{\rm mix}^{(LL)(RR)}|$ (right)
      in $\tau^- \to \mu^- \mu^- \mu^+$. \label{plotmix}
}
\end{figure}

Finally, we have to take into account the contributions from the interference terms
between the radiative operators and four-fermion operators, for the cases where only
the chirality of the $\tau$\/-lepton has to be flipped.
The interference term between (\ref{LLLL}) and (\ref{rad}) reads
\begin{equation}
\frac{\text{d}^2\Gamma_{\rm mix}^{(LL)(LL)}}{\text{d}m_{23}^2 \text{d}m_{12}^2} 
=
\alpha_\text{em} \,
\frac{ 2  \, v \, \text{Re}[ g_{\rm V}^{(L_\mu L^\tau)(L_\mu L^\mu)} 
\, g_{\rm rad}^{*(L_\mu R^\tau)} ] }{\Lambda^4}  
\left[
\frac{m_{12}^2-3 m_\mu^2}{64 \, \pi^3 \, m_\tau^2} 
+ \frac{m_\mu^2 (m_\tau^2-m_\mu^2)(m_{13}^2+m_{23}^2)}
       {128 \, \pi^3 \, m_\tau^2 \, m_{13}^2 \, m_{23}^2}
\right]
\,,
\label{dMix1}
\end{equation}
The interference between (\ref{LLRR}) and (\ref{rad}) results in
\begin{equation}
\begin{split}
\frac{\text{d}^2\Gamma_{\rm mix}^{(LL)(R R)}}{\text{d}m_{23}^2 \text{d}m_{12}^2}
 = \ &
\alpha_\text{em} \,
\frac{ 2 \, v \, \text{Re}[ g_{\rm V}^{(L_\mu L^\tau)(R_\mu R^\mu)} \, g_{\rm rad}^{*(L_\mu R^\tau)} ]}{\Lambda^4} 
\\[0.1em]
& \quad {} \times
\left[
\frac{m_\tau^2-m_{12}^2-3 m_\mu^2}{256 \, \pi^3 \, m_\tau^2} 
+ \frac{m_\mu^2 (m_\tau^2-m_\mu^2)(m_{13}^2+m_{23}^2)}
       {256 \, \pi^3 \, m_\tau^2 \, m_{13}^2 \, m_{23}^2}
\right] 
\,.
\label{dMix2}
\end{split}
\end{equation}
In both cases, the photon pole at $m_{13}^2=0$ or $m_{23}^2=0$ is suppressed by
the small muon mass. The remaining terms increase (decrease) monotonically 
with $m_{12}^2$, respectively, see Fig.~\ref{plotmix}.

Combining (\ref{dGam1}-\ref{dMix2}) and integrating over phase space, we obtain
for the total decay width (normalized to the SM decay $\tau \to \mu \bar \nu_\mu \nu_\tau$,
and neglecting the muon mass)
\begin{eqnarray}
\begin{aligned}
  \frac{\Gamma[\tau^- \to \mu^-\mu^-\mu^+]}{\Gamma[\tau^-\to \mu^- \bar\nu_\mu \nu_\tau]}
\, = \, \frac{1}{G_F^2 \Lambda^4} & \left\{ \frac{2\,| g_V^{(L L)(L L)}|^2+ 2\,| g_V^{(R R)(R R)}|^2 + |g_V^{(L L)(R R)}|^2 + |g_V^{(R R)(L L)}|^2}{8}
\right.
\\ & \left. + \frac{ \alpha_{\rm em}^2 \, v^2}{m_\tau^2} \left( \ln \frac{m_\tau^2}{m_\mu^2} - \frac{11}{4} \right) \left(|g_{\rm rad}^{(LR)}|^2 + |g_{\rm rad}^{(RL)}|^2 \right) 
\right.
\\ & \left.
+ \frac{ \alpha_{\rm em} \, v }{ 2 m_\tau} 
  {\rm Re}\left[ 
  2 \, g_{\rm rad}^{*(LR)} \, g_V^{(LL)(LL)} 
 +  g_{\rm rad}^{*(LR)} \, g_V^{(LL)(RR)} + (L \leftrightarrow R)
\right]
\right\}\,.
\end{aligned}
\end{eqnarray}
This result is consistent with the formula quoted, for instance, in \cite{Brignole:2004ah}.

\subsection{The Decay \boldmath $\tau^- \to e^- \mu^- \mu^+$ \boldmath}

\label{sec:taue}

For completeness, we also discuss the Dalitz distributions for the decay mode
$\tau^- \to e^- \mu^- \mu^+$, which belongs to the class of decays with three
\emph{different} particles in the final state. Again, we will give the results in
terms of the invariant masses
\begin{equation}
m_{--}^2 \equiv m_{12}^2 = (p_{e^-} + p'_{\mu^-})^2 \,, \qquad
m_{+-}^2 \equiv m_{23}^2 = (p'_{\mu^-} + p_{\mu^+})^2 \,, 
\end{equation}
and $m_{13}^2 = m_\tau^2+2m_\mu^2 - m_{--}^2-m_{+-}^2$,
where we set the electron mass to zero.\footnote{Notice
that the photon pole from $\tau^- \to e^-\gamma^* \to e^- \mu^- \mu^+$ is still
regulated by the muon mass.}

From the purely left-handed term in the effective Hamiltonian,
$H_{\rm eff}^{(LL)(LL)}$ in (\ref{LLLL}), we obtain
the Dalitz distribution
\begin{equation}
\frac{\text{d}^2\Gamma_V^{(LL)(LL)}}{\text{d}m_{23}^2 \, \text{d}m_{12}^2} 
=
\frac{| g_V^{(L_e L^\tau )(L_\mu L^\mu)}|^2}{\Lambda^4} \, 
\frac{m_\tau^4 - (2 m_{12}^2 -m_\tau^2-2m_\mu^2)^2}{512 \, \pi ^3\, m_{\tau }^3} \,,
\label{dGam1e}
\end{equation}
which (except in the vicinity of the phase-space boundaries)
coincides with (\ref{dGam1}) up to corrections of order $m_\mu^2/m_\tau^2$
and a statistical factor.
Consequently, the corresponding Dalitz plot looks almost identical to Fig.~\ref{plot} (left).

\begin{figure}[tp]
\begin{center}
 \begin{minipage}{0.48\textwidth}
       \centering
       \centering
        \includegraphics[width=0.78\textwidth]{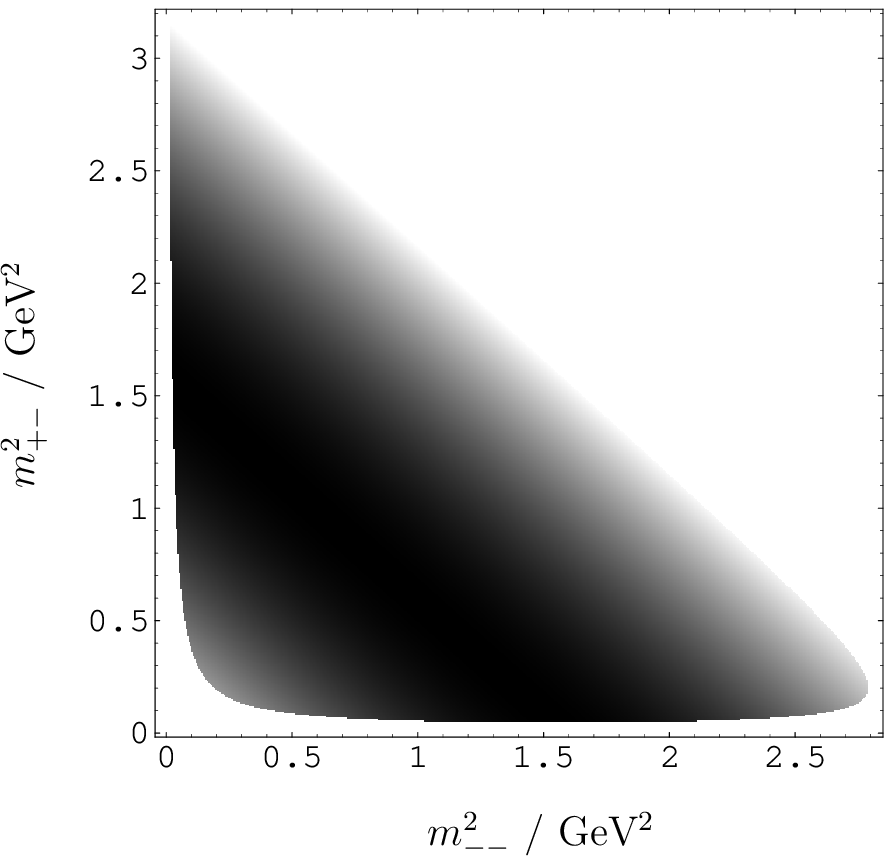}
\end{minipage} \hfill
\begin{minipage}{0.48\textwidth}
       \centering
       \centering
        \includegraphics[width=0.78\textwidth]{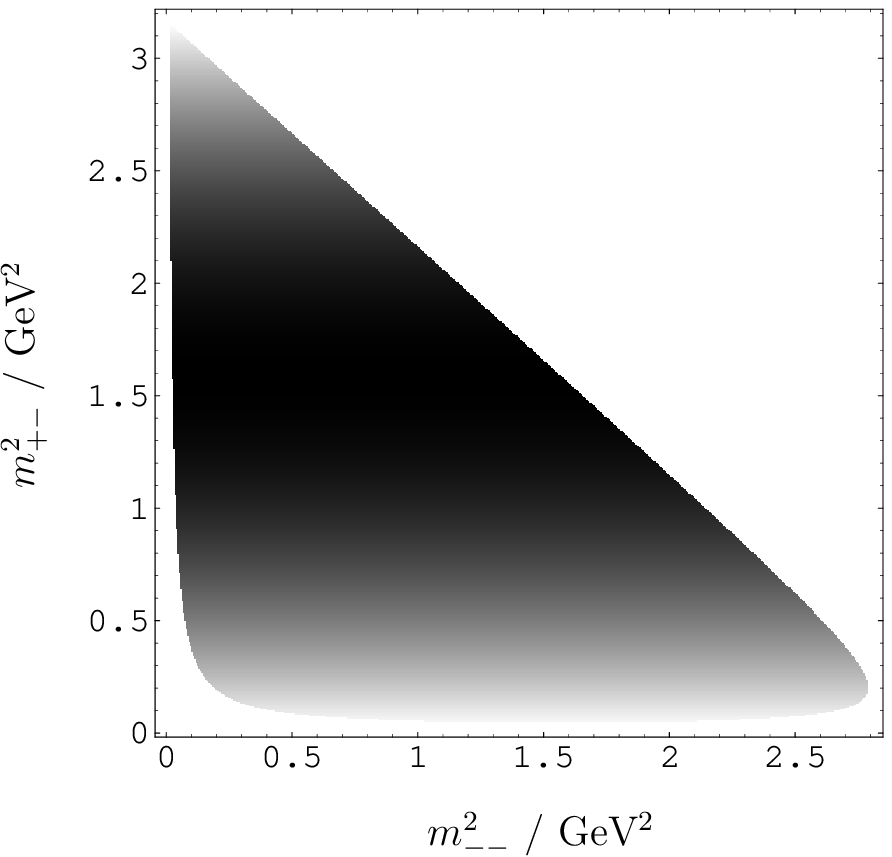}
\end{minipage}
\end{center}
  \caption{Dalitz plots for the two contributions to $\text{d}^2\Gamma_V^{(LL)(RR)}$
      in $\tau^- \to e^- \mu^- \mu^+$. \label{plotLLRRe}}
\end{figure}

From $H_{\rm eff}^{(LL)(RR)}$ in (\ref{LLRR}) we obtain
for the case of a left-handed $\tau$\/-lepton
\begin{equation}
\begin{split}
 \frac{\text{d}^2\Gamma_V^{(LL)(RR)}}{\text{d}m_{23}^2 \text{d}m_{12}^2} 
= \ & 
\frac{| g_V^{(L_e L^\tau)(R_\mu R^\mu)}|^2}{\Lambda^4}
\,
\frac{
m_\tau^4 
- ( 2 m_{13}^2 - m_\tau^2- 2 m_\mu^2)^2}
{512 \, \pi^3\, m_{\tau }^3}
\\[0.2em]
& {} 
+\frac{| g_V^{(L_\mu L^\tau)(R_e R^\mu)}|^2}{\Lambda^4}
\,
\frac{(m_\tau^2-2m_\mu^2)^2 -
(2 m_{23}^2 - m_\tau^2 - 2 m_\mu^2)^2}
{512 \, \pi^3\, m_{\tau }^3}
\,.
\end{split}
\label{dGam2e}
\end{equation}
The corresponding Dalitz plots for the two separate contributions are
shown in Fig.~\ref{plotLLRRe} (left: the term $\propto | g_V^{(L_e L^\tau)(R_\mu R^\mu)}|^2$; right: the term $\propto| g_V^{(L_\mu L^\tau)(R_e R^\mu)}|^2$).
The events are distributed around $m_{--}^2 + m_{+-}^2 \simeq m_\tau^2/2$
or $m_{+-}^2 \simeq m_\tau^2/2$, respectively.
For equal coupling constants in (\ref{dGam2e}) we recover the $\tau \to 3\mu$
case in (\ref{dGam2}) (again up to mass corrections and a statistical factor).

\begin{figure}[h!tp]
       \centering
        \includegraphics[width=0.39\textwidth]{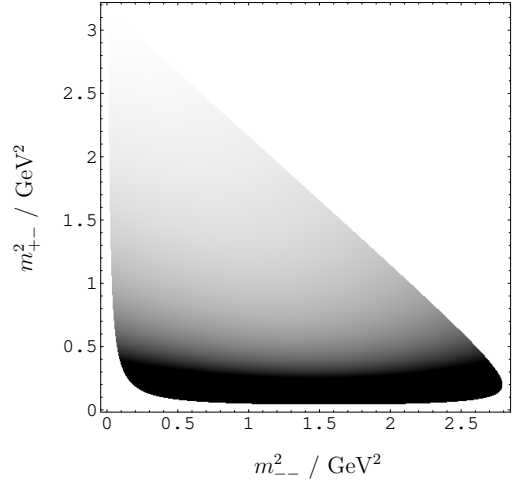}
        \caption{Dalitz plot for $\text{d}^2\Gamma_{\rm rad}^{(LR)}$ 
      in $\tau^- \to e^- \mu^- \mu^+$. \label{plotrade}}
\end{figure}

For the radiative decay operators, we obtain
\begin{equation}
\begin{split}
\frac{\text{d}^2\Gamma_{\rm rad}^{(LR)}}{\text{d}m_{23}^2 \text{d}m_{12}^2} 
=
& \,
\alpha_{\text{em}}^2 \, 
\frac{|g_{\rm rad}^{(L_e R^\tau)}|^2 \, v^2}{\Lambda^4}
 \left[
\frac{m_\mu^2 \, (m_{23}^2-m_\tau^2)^2}{64 \, \pi^3 \, m_\tau^3 \,m_{23}^4}
+
\frac{m_{12}^4+m_{13}^4-2 m_\mu^4}{128 \, \pi^3 \, m_{\tau }^3 \, m_{23}^2}
+ \frac{ m_\tau^2-m_{23}^2}{128 \, \pi^3 \, m_{\tau }^3}\right]
\,,
\end{split}
\label{dGam3e}
\end{equation}
and the corresponding Dalitz plot is shown in Fig:~\ref{plotrade}. In this
case the photon pole enhances the events at low values of $m_{+-}^2=m_{23}^2$.

\begin{figure}[h!bp]
      \begin{minipage}{0.48\textwidth}
       \centering
        \includegraphics[width=0.78\textwidth]{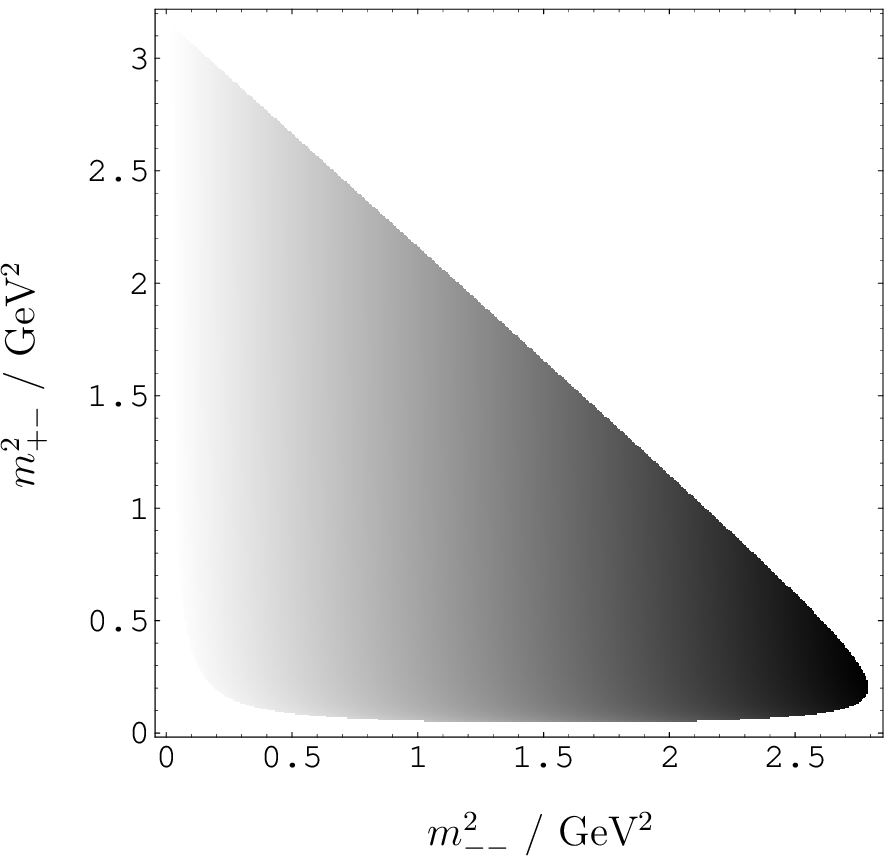}
      \end{minipage}\hfill
      \begin{minipage}{0.48\textwidth}
       \centering
        \includegraphics[width=0.78\textwidth]{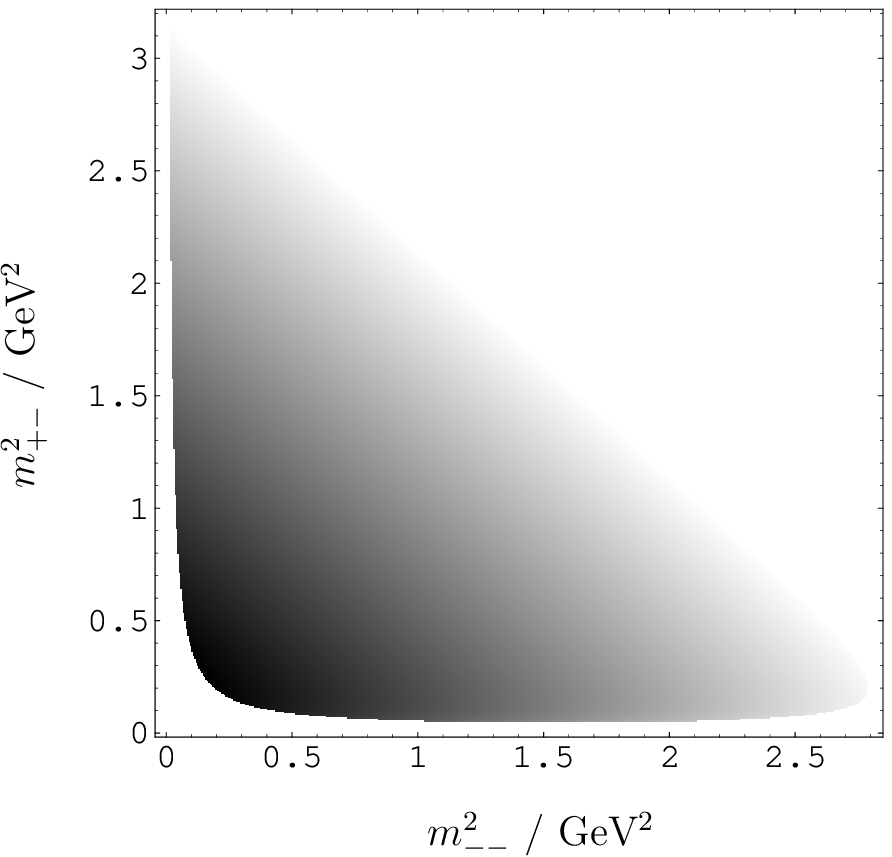}
      \end{minipage}
\caption{Dalitz plot for $|\text{d}^2\Gamma_{\rm mix}^{(LL)(LL)}|$ (left)
and $|\text{d}^2\Gamma_{\rm mix}^{(LL)(RR)}|$ (right)
      in $\tau^- \to e^- \mu^- \mu^+$. \label{plotmixe}}
\end{figure}

Finally, for the interference terms between (\ref{LLLL}) and (\ref{rad}) 
we get
\begin{equation}
\frac{\text{d}^2\Gamma_{\rm mix}^{(LL)(LL)}}{\text{d}m_{23}^2 \text{d}m_{12}^2} 
=
\alpha_\text{em} \,
\frac{ 2 \, v \, \text{Re}[ g_{\rm V}^{(L_e L^\tau)(L_\mu L^\mu)} \, g_{\rm rad}^{*(L_e R^\tau)} ]}{\Lambda^4}  \,
\left[
\frac{m_{12}^2-2 m_{\mu}^2}{128 \pi ^3 m_{\tau }^2} 
+
\frac{m_{\mu }^2}{128 \pi ^3 m_{23}^2} 
\right] 
\,,
\label{dMix1e}
\end{equation}
and 
\begin{equation}
\frac{\text{d}^2\Gamma_{\rm mix}^{(LL)(RR)}}{\text{d}m_{23}^2 \text{d}m_{12}^2} 
=
\alpha_\text{em} \,
\frac{ 2 \, v \, \text{Re}[ g_{\rm V}^{(L_eL^\tau)(R_\mu R^\mu)} \, 
 g_{\rm rad}^{*(L_e R^\tau)} ]}{\Lambda^4}  \,
\left[
\frac{m_{13}^2-2 m_\mu^2}{128 \pi ^3 m_{\tau }^2} 
+
\frac{m_{\mu }^2}{128 \pi ^3 m_{23}^2} 
\right]
\,.
\label{dMix2e}
\end{equation}
The corresponding Dalitz plots are shown in Fig.~\ref{plotmixe}.

\clearpage

\clearpage

%%%%%%%%%%%%%%%%%%%%%%%%%%%%%%%%%%%%%%%%%%%%%%%%%%%%%%%%%%%%%%%%%%%%%%%%%%%%%%%%%%%%%%%%%%%%%%

\section{Discussion and Conclusions}
\label{sec:concl}

Lepton-number violating processes, like the decay
$\tau^- \to \ell \ell' \ell''$ discussed in this paper,
provide an important test of the Standard Model. 
Since many new physics models allow for dramatic enhancements 
compared to the tiny SM effects, there is a potential to
falsify the SM by measuring such decays at future experiments. 
At the same time, the foreseen improvement of experimental limits may further 
tighten the constraints on specific new physics models.

In both cases, the event distributions in phase space provided in this work, will be
helpful. The difference between the Dalitz distributions arising 
from four-lepton operators, which show a rather uniform behaviour, 
and the distributions from radiative operators
which are concentrated at small values of $m_{+-}^2=(p_{\ell^+}+p_{\ell^-})^2$,
clearly gives a handle to disentangle different new physics models already 
on the basis of rather few events. Typically, these models give
rather different predictions for the relative size of radiative and
four-fermion operators, as we will discuss in the following.

In super-symmetric extensions of the SM, one typically finds that the
photon-dipole operator, induced by penguin diagrams,
dominates over the four-lepton operators. This leads to simple 
correlations like (see e.g.\ \cite{Brignole:2004ah})
\[
 \frac{\Gamma(\tau\to 3\mu)}{\Gamma(\tau \to \mu\gamma)}
\simeq
 \frac{\alpha_{\rm em}}{3\pi} \left( \ln \frac{m_\tau^2}{m_\mu^2} - \frac{11}{4} \right)
= {\cal O}(10^{-3})
\,.
\]
In this case, one  expects Dalitz distributions as shown in Fig.~\ref{plotrad}.

It has been pointed out in \cite{Paradisi:2005fk} that Higgs-mediated $\tau \to \mu$
transitions may alter this result, if $\tan\beta$ and the off-diagonal
slepton mass-matrix element $\delta_{3\ell}$ are large. 
For instance, in the decoupling limit  ($\cos(\beta-\alpha) =0$, $m_{A^0} \gg M_Z$) the author
of \cite{Paradisi:2005fk} finds
\[
\frac{\Gamma(\tau \to \ell \mu \mu)}{\Gamma(\tau \to \ell \gamma)}
\le \frac{3 + 5 \delta_{\ell \mu}}{36} \sim {\cal O} (0.1) 
\] 
where $\delta_{\ell \mu} = \tilde{m}^2_{\ell \mu} / \tilde{m}^2$.
Testing such scenarios in experiment will be more involved, as one
generally has to allow for the interplay of all contributions to the
Dalitz distributions, Fig.~\ref{plot}-\ref{plotmix}.

The situation is somewhat different in the case of Little Higgs Models with T-Parity 
(LHT) \cite{Goyal:2006vq,Choudhury:2006sq,BurasBlanke}. Here 
the $Z_0$ and box-diagram contributions 
dominate compared to the radiative operators \cite{BurasBlanke}, which is mainly due to the constructive (respectively
destructive) interference between the individual heavy gauge boson contributions. 
Depending on the parameter values of the LHT, one finds
\[
\frac{\Gamma(\tau \to 3\mu)}{ \Gamma(\tau \to \mu \gamma)}
= {\cal O}(1)
\] 
for a mass scale of the LHT mirror fermions of about 1~TeV.
In this case, one can expect rates for LFV decays which are already close
to the present bounds. Because of the sub-dominance of the radiative
dipole operator, we expect a rather flat Dalitz distribution for
$\tau \to 3\mu$, as illustrated in Fig.~\ref{plot}.

\subsection*{Acknowledgements}
This work was supported by the German Research Foundation DFG \\ under 
contract No.~MA1187/10-1.

\end{document}